\documentclass{article}
\usepackage{glas}
\usepackage{epsfig}
\begin{document}

\def\lapproxeq{\lower .7ex\hbox{$\;\stackrel{\textstyle <}{\sim}\;$}}
\def\gapproxeq{\lower .7ex\hbox{$\;\stackrel{\textstyle >}{\sim}\;$}}
\newcommand{\ee}{\mbox{$e^+e^-$}}
\newcommand{\logxp}     {\mbox{$\ln(1/x_{p})$}}

\begin{titlepage}{GLAS-PPE/1999--01}{January 1999}
\title{
Momentum Spectra in the\\
Current Region of the Breit Frame\\
in Deep Inelastic Scattering \\
at HERA }
\author{N.\ Brook} 
\collaboration{on behalf of the ZEUS collaboration}

\begin{abstract}
The production of charged particles 
has been measured in  deep
inelastic scattering (DIS) with the ZEUS detector.
The evolution of the moments of the
scaled momenta distributions in $Q^2$ and $x$
has been
investigated in the current fragmentation region of the
Breit
frame. The results in the current region are compared to
$e^+e^-$ data and  QCD analytical calculations.
The results are consistent
with the universality of
single-particle spectra  in DIS and  $e^+e^-$ annihilation at
high $Q^2.$
The results at low $Q^2$
disagree with analytical calculations based on the 
modified leading log approximation (MLLA) 
and local parton hadron duality (LPHD).
\end{abstract}

\vfill
\conference{talk given at the \\
 3rd UK Phenomenology Workshop \\
on HERA Physics, \\
Durham, United Kingdom. \\Sept 20-25, 1998}
\end{titlepage}

\section*{Introduction}

This paper reports the results of a study
of the properties of the hadronic final
state in positron-proton deep inelastic scattering (DIS). 
The event kinematics of DIS are determined by the negative square of the
four-momentum transfer of the virtual exchanged boson,
$Q^2\equiv-q^2$, and the Bjorken scaling variable, $x=Q^2/2P\!\cdot\!q$,
where $P$ is the four-momentum of the proton.
In the Quark Parton Model (QPM),
the interacting quark from the proton carries four-momentum $xP.$
The variable $y$, the fractional energy transfer to the proton in its
rest frame, is related to $x$ and $Q^2$ by $y\approx Q^2/xs$,
where $\sqrt s$ is the positron-proton centre of mass energy.

A natural frame in which to study the dynamics of the hadronic final
state
in DIS is the Breit frame~\cite{feyn}.
In this frame the exchanged
virtual boson is completely space-like and has a four-momentum
$q = (0,0,0,-Q=-2xP^{Breit})\equiv (E,~p_x,~p_y,~p_z)$,
where $P^{Breit}$ is the momentum of the proton in the Breit frame.
The particles produced in the
interaction can be assigned to one of two regions:
the current region if
their $z$-momentum in the Breit frame is negative, and
the target region if their $z$-momentum is positive.
The advantage of this
frame is that it gives a
maximal  separation of the incoming and outgoing partons
in the QPM. In this model
the maximum momentum a particle can have in the current region
is $Q/2.$

The current region in the Breit frame
is analogous to a single hemisphere of $e^+e^-$ annihilation.
In $e^+e^- \rightarrow q \bar q$ annihilation the two quarks are produced
with equal and opposite momenta, $\pm \sqrt{s}/2.$
The fragmentation of these quarks can be compared with
that of the quark struck from the
proton which has outgoing momentum $-Q/2$ in the Breit frame.
In the direction of this struck quark
the scaled momentum spectra of the particles, expressed in terms of
$x_p = 2p^{Breit}/Q,$
are expected to have a
dependence on $Q$ similar to that observed 
in \ee~annihilation~\cite{eedis,anis,char} at energy $\sqrt{s}=Q.$

Within the modified leading log approximation (MLLA)
there are predictions of how the higher order
moments of the parton momentum spectra should evolve
with energy scale~\cite{fongweb,dokevol}.
The parton level predictions depend on
 two free parameters, a running
strong coupling, governed by a QCD scale
$\Lambda,$ and an energy cut-off, $Q_0,$
below which the parton
evolution is truncated.
The hypothesis of
local parton hadron duality (LPHD)~\cite{LPHD},
which relates the observed hadron
distributions to the calculated parton distributions via a constant of
proportionality, is used in conjunction with the predictions
of the MLLA allowing the calculation to be directly compared with data.

\section*{Results}

The moments of the \logxp\ distributions have been investigated
up to the 4th order;
the mean $(l),$ width $(w),$ skewness $(s)$ and kurtosis $(k)$
 were extracted from the distribution
by fitting a distorted Gaussian of the following form:
\begin{displaymath}
\frac{1}{\sigma_{tot}} \frac{d\sigma}{d\ln(1/x_p)} \propto
 \exp\left(\frac{1}{8}k-\frac{1}{2}s\delta -\frac{1}{4}(2+k)\delta^2
+\frac{1}{6}s\delta^3 + \frac{1}{24}k\delta^4\right) \label{eq:dg}
\end{displaymath}
where $\delta = (\ln(1/x_p) - l)/w,$
over a range of 3 units ($Q^2 < 160 {\rm\ GeV^2}$) or
4 ($Q^2 \ge 160 {\rm\ GeV^2}$) units in $\ln(1/x_p)$ around the mean.
The equation
 was motivated by the expression used for the MLLA predictions of the spectra
in ref.~\cite{fongweb}.

\begin{center}
\begin{figure}[hbt]
\centering
\centerline{\psfig{figure=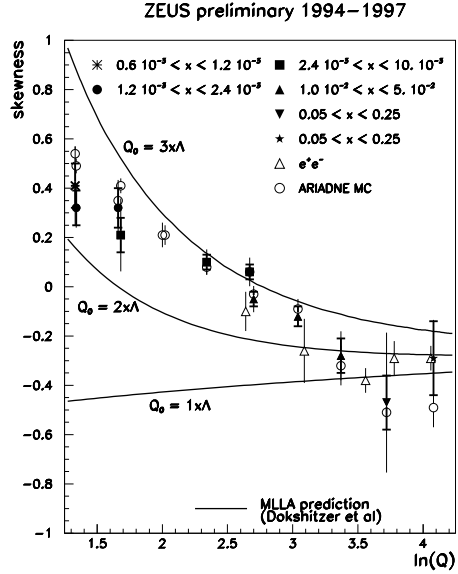,height=8.cm}}
\caption{ Evolution of the skewness of the
$\ln(1/x_p)$ distribution with $\ln(Q).$
Data from $e^+e^-$ and $ep$ are shown together with, ARIADNE
Monte Carlo predictions and 
the MLLA predictions of
Dokshitzer {\it et al}~\protect\cite{dokevol}
predictions (the full line is $Q_0=\Lambda,$ the dashed
$ Q_0= 2\Lambda,$ and the dotted $Q_0 = 3\Lambda$).
The overlapping points are different $x$ ranges in the same $Q$ range.
The
inner error bars are the statistical error and the outer error bars are the
systematic and statistical errors added in quadrature.}
\label{fig:qevol}
\end{figure}
\end{center}

Figure~\ref{fig:qevol} shows the skewness of the  $\ln(1/x_p)$
spectra as a function of $\ln(Q).$
It is evident that the skewness decreases with increasing $Q.$
Similar fits performed on \ee\ data 
shows a reasonable agreement with our results at high $Q^2,$
consistent with the universality of fragmentation for this distribution.
The ARIADNE Monte Carlo model~\cite{ariadne}
 gives a reasonable description of the data.
The data are compared with the MLLA predictions of ref.~\cite{dokevol},
using a value of $\Lambda=175{\rm \ MeV},$
for different values of $Q_0.$ 
The MLLA calculations predict a
negative skewness which decreases towards zero
with increasing $Q$ in the case of the limiting spectra
($Q_0 = \Lambda$).
This is contrary to the measurements. A reasonable description of the
behaviour of the skewness with $Q$ can be achieved for a
truncated cascade ($Q_0 > \Lambda$), but a consistent description of
the mean, width, skewness and kurtosis
cannot be achieved~\cite{newzeus}.
A range of $\Lambda$ values were investigated and none
gave a good description of all the moments.
We conclude that the
MLLA predictions, assuming LPHD, do not describe the data.
It should be noted though that  a moments analysis has been
performed~\cite{seroch}, taking into
account the limitations of the  massless assumptions of the MLLA predictions,
where good agreement was found between the
limiting case of the MLLA
and ${\rm e^+e^-}$ data over a large range of energy,
$ 3.0 < \sqrt{s} < 133.0\ {\rm GeV}.$

\section*{Summary}

Charged particle distributions have been studied in the current region
of the Breit frame in DIS. 
The moments of the $\ln(1/x_p)$ spectra in the current region
at high $Q^2$
exhibit the same energy scale behaviour as that observed
in $e^+e^-$ data.
The moments cannot be described by
the MLLA calculations together with LPHD.

\end{document}